\documentclass[epj]{svjour}
\usepackage{epsfig} \usepackage{psfig} \usepackage{times}

\begin{document}
\title{In-medium spectral functions of charmonia  studied by
 $\bar p + A$ reactions \footnote{supported by DFG, RFFI and
Forschungzentrum J\"{u}lich.}}
\author{ Ye. S. Golubeva\inst{1}, E. L. Bratkovskaya\inst{2},
W. Cassing\inst{3}, L. A. Kondratyuk\inst{4}} \institute{Institute
for Nuclear Research, 60th October Anniversary Prospect 7A,
117312 Moscow, Russia  \and Institut f\"ur Theoretische Physik,
Universit\"at Frankfurt, Robert-Mayer Str. 8-10,  D-60054 Frankfurt am Main, Germany
\and Institut f\"ur Theoretische Physik, Universit\"at Giessen, Heinrich-Buff-Ring 16,
D-35392 Giessen, Germany   \and Institute of Theoretical and
Experimental Physics, B.\ Cheremushkinskaya 25,
 117259 Moscow, Russia}
\date{Received: date / Revised version: date}

\abstract{We study the perspectives of resonant charmonium
production in $\bar{p} + A$ reactions within the Multiple
Scattering Monte Carlo (MSMC) approach. We calculate the
production of the resonances $\Psi(2S)$ and $\Psi(3770)$ on
various nuclei, their propagation and decay to dileptons and $D+
\bar{D}$ in the medium and vacuum, respectively, employing
parametrizations for the $D,\bar{D}$ self energies taken from QCD
sum rule studies.  The elastic and inelastic interactions of the
charmonia and open charm mesons in the medium are taken into
account, too. It is found that  the $D, \bar{D}$ invariant mass
spectra from light and heavy nuclei are not sufficiently sensitive
to the in-medium properties of the $\Psi(2S)$ and $\Psi(3770)$.
However, a 'suppression' of low mass dileptons from the
$\Psi(3770)$ might be seen experimentally as well as a small
broadening of the $\Psi(2S)$ dilepton spectra. }


\authorrunning{Ye. S. Golubeva et al.} \titlerunning{Interactions of
charmed mesons with nucleons in the $\bar p A$ reaction}

\maketitle

\section{Introduction}
The dynamics of charmonia and open charm mesons at finite baryon
density and/or temperature  has gained sizeable interest
especially in the context of a phase transition from hadronic
matter to a quark-gluon plasma (QGP), where charmed meson states
should no longer be formed due to color screening
\cite{Satz,Satz2}. However, the suppression of the $c\bar{c}$
vector mesons in the high density phase of nucleus-nucleus
collisions as seen by the NA50 Collaboration \cite{NA50} might
also be attributed to inelastic comover scattering (cf.
\cite{Capella,Capella2,Cass97a,Cass99,Vogt99,Gerschel} and Refs.
therein) if the corresponding charmonium-hadron cross sections are
in the order of a few mb \cite{Haglin,Oh,I1,I2,I3,Ko}.
Theoretical models here differ in their predictions by at least an
order of magnitude \cite{Bernd} such that at present no stringent
conclusions can be drawn. As pointed out by Seth \cite{Seth} this
uncertainty can be substantially reduced by experimental data from
$\bar{p}$ induced reactions on nuclei, where charmonia can be
formed on resonance with moderate momenta relative to the target
nucleus. Such studies might be performed experimentally at the
HESR, which is proposed as a future facility at GSI-Darmstadt
\cite{GSIfut}.

In previous studies we have explored the perspectives of charmed
meson - nucleon scattering in the $\bar{p} d$ reaction
\cite{Cass00}, where the charmonium is produced on resonance with
the proton in the deuteron and may scatter with the spectator
neutron. We have also investigated the production  of the resonances $\Psi(3770)$,
$\Psi(4040)$ and $\Psi(4160)$ on various nuclei, their propagation
and decay to $D, \bar{D}, D^*, \bar{D}^*, D_s, \bar{D}_s$ in the
medium and vacuum \cite{Golub02}, respectively. Furthermore, the
elastic and inelastic interactions of the charmonia and open charm
mesons in the medium have been followed up in the Multiple
Scattering Monte Carlo   approach (MSMC) to study the collisional
effects as a function of the target mass $A$ \cite{Golub02}. In
our previous studies we have discarded the real part of the meson
self energies $\Re \Pi(M)$; the latter can be expressed as
single-particle potentials or 'mass shifts' via $\Re \Pi/(2 M)$
and might be calculated by dispersion relations.

Another way of getting information about mass shifts of hadrons in
the medium are QCD sum rules. Related studies for $J/\Psi$ mesons
suggest that the mass shift $\Re \Pi_V/(2 M_V)$ at normal nuclear
matter density is only in the order of a few MeV
\cite{Klingl,Haya2} due to a small coupling of the $c, \bar{c}$
quarks to the nuclear medium. We thus will assume $\Re \Pi_V
\approx$  0 ($V = J/\Psi, \Psi(2S), \Psi(3770))$ throughout the
following study. However, the open charm mesons ($D, \bar{D}, D^*,
\bar{D}^*$)  are expected to show sizeable mass shifts due to
their light quark content \cite{Alex99,Sibirtsev,Haya,Tsu2} in
analogy to the kaons and antikaons. Especially for dropping
$D,\bar{D}$ effective masses in the medium the decay width of the
vector mesons then should be enhanced at finite density $\rho_A$
as suggested in Ref.
\cite{Haya}.

We recall, that experiments on $K^\pm$ production from
nucleus-nucleus collisions at SIS energies of 1-- 2 A$\cdot$ GeV
have shown that in-medium properties of the kaons are seen in the
collective flow pattern of $K^+$ mesons both, in-plane and
out-of-plane, as well as in the abundancy  and spectra of
antikaons \cite{Cass99,Kaos1,Kaos2,KaoS3,cmko,Li2001}. In-medium
modifications of the mesons have become a topic of substantial
interest in the last decade triggered in part by the early
suggestion of Brown and Rho \cite{BR,BR2}, that the modifications
of hadron masses should scale with the scalar quark condensate
$<q\bar{q}>$ at finite baryon density. Thus, in view of the
analogy to the $K^\pm$ mesons -- exchanging $s, \bar{s}$ by $c,
\bar{c}$ quarks -- in-medium modifications of the open charm
mesons $D, \bar{D}$ should be observable, too. First exploratory
studies by Sibirtsev et al. \cite{Alex99} have suggested also an
enhanced production of $D, \bar{D}$ mesons in $\bar{p} + A$
reactions at subthreshold energies. Such enhanced cross sections
should even more prominantly be seen in central $Au+Au$ collisions
at 20 to 25 A$\cdot$GeV as advocated in Ref. \cite{Cass01}.
However, some note of caution appears necessary since enhanced
cross sections might arise due to secondary reactions channels
\cite{Cass96,Aichelin} and thus do not provide a unique signal for
in-medium mass shifts. It is the aim of the present work to
explore the possibility, if such mass shifts for open charm mesons
might be extracted directly from dilepton or $D,\bar{D}$ invariant
mass spectroscopy.

Our work is organized as follows: In Section 2 we will briefly
describe the input cross sections to the MSMC approach and
evaluate the formation cross sections for the resonances
$\Psi(2S)$ and $\Psi(3770)$ on protons and nuclei. The fractional
decay of these resonances to open charm mesons and dileptons in
the medium and vacuum, respectively, is calculated in Section 3
using 'free' and 'in-medium' spectral functions. Section 4
concludes this paper with a summary and discussion of open
problems.

\section{Resonance production and decay in $\bar{p} A$ reactions}
We here examine the possibility to measure the in-medium life time
(or total width) of the resonances $V=(\Psi(2S),\Psi(3770)$)
produced in $\bar{p} A$ reactions.  To this aim we describe the
vector meson spectral function by a Breit-Wigner distribution
\begin{equation}
A_V(M^2) = N_V \frac{M
\Gamma_{tot}^V(M)}{(M^2-M_V^2-\Re\Pi_V(M))^2 + M^2
\Gamma_{tot}^2(M)}, \label{BW}
\end{equation}
where $M_V$ denotes the vacuum pole mass, $\Re\Pi_V$ is the vector
meson self energy in the medium - which vanishes in the vacuum -
and $N_V \sim 1/\pi$ (for small width $\Gamma_{tot}$) is a
normalization factor that ensures $\int dM^2 A(M^2) =1$. The total
width $\Gamma_{tot}$ is separated into decay and collisional
contributions as
\begin{equation}
 \Gamma_{tot}(M,\rho_A) =  \Gamma_{dec}(M,\rho_A) +
\Gamma_{coll}(M,\rho_A), \label{gamma}
\end{equation}
where \begin{equation} \Gamma_{dec}(M,\rho_A) = \sum_c
\Gamma_c(M,\rho_A) \label{vac}
\end{equation} with $\Gamma_c(M,\rho_A)$ denoting the partial
width to the decay channels $c \equiv D\bar{D}, D\bar{D}^*, D^*
\bar{D}$, $e^+ e^-, \mu^+ \mu^- $ etc.,  respectively. If the $D,
\bar{D}$ meson properties change with density $\rho_A$, this will
modify the total and partial widths accordingly. The partial
widths for electromagnetic decays, furthermore, are assumed to be
independent on the nuclear medium.

The in-medium collisional width $\Gamma_{coll}(M,\rho_A)$ in
(\ref{gamma}) is determined from the imaginary part of the forward
vector-meson nucleon scattering amplitude as
\begin{equation}
\Gamma_{coll} = \frac{4 \pi}{M_V} \Im f_V(0) \rho_A, \label{coll}
\end{equation}
where $\rho_A$ again denotes the nuclear density. Furthermore,
$\Im f_V(0)$ is determined via the optical theorem by the total
vector-meson nucleon cross section $\sigma_{VN}$ that will be
specified below.

\subsection{In-medium properties of open charm mesons}
Whereas the pole masses and widths of the vector mesons $M_V$ and
open charm mesons in vacuum are rather well known \cite{PDG} we
have to model the in-medium properties  that enter the spectral
function (\ref{BW}) via (2) and (\ref{vac}). As noted before,
there should be also shifts of the vector meson pole masses by
$\Re \Pi_V(M)$. According to QCD sum rule results \cite{Klingl}
the $J/\Psi$ meson mass shift is only in the order of a few MeV at
nuclear matter density $\rho_0$ and thus of minor importance.
However, QCD sum rules have limited predictive power for the
excited states $\Psi(2S)$ and $\Psi(3770)$ and sizeable mass
shifts might occur e.g. from $D \bar{D}$ loops in case of the
$\Psi(3770)$ \cite{Kolee}. Here we do not speculate on such
effects and will discard explicit mass shifts of the $\Psi(2S)$
and $\Psi(3770)$ and concentrate on the modification of the decay
channels in the medium. To this end in each channel $c$ the
relative decay to mesons $m_1$ and $m_2$ is described by a matrix
element (squared) and phase space, i.e.
\begin{equation}
\Gamma_c \sim |{\cal M}_c|^2 p_c^3, \label{channel}
\end{equation}
where $p_c$ is the meson momentum in the rest frame of the
resonance. The assumption (\ref{channel}) should hold in leading
order, however, might fail if the vector meson has an explicit
internal structure as advocated e.g. in Ref. \cite{Friman}.
Nevertheless, in case of the $\Psi(3770)$ only the lowest channel
to $D \bar{D}$ contributes such that the matrix element in
(\ref{channel}) can be fixed by the total width at half maximum in
vacuum, which we take as $\Gamma_{\rm{FWHM}} =$ 25 MeV \cite{PDG}.
We assume the same matrix element also for the $\Psi(2S)$, which
cannot be controlled experimentally since this state (of mass
3.686 GeV) is below the $D+\bar{D}$ threshold. The resulting
spectral functions $S(M)=2 M A(M)$ -- including the
electromagnetic and light hadron decay widths -- are shown
in Fig. 1 for the
$\Psi(2S)$ (upper part) and $ \Psi(3770)$ (lower part) for the 'free' case by the solid lines.

\begin{figure}[t]
\centerline{\psfig{figure=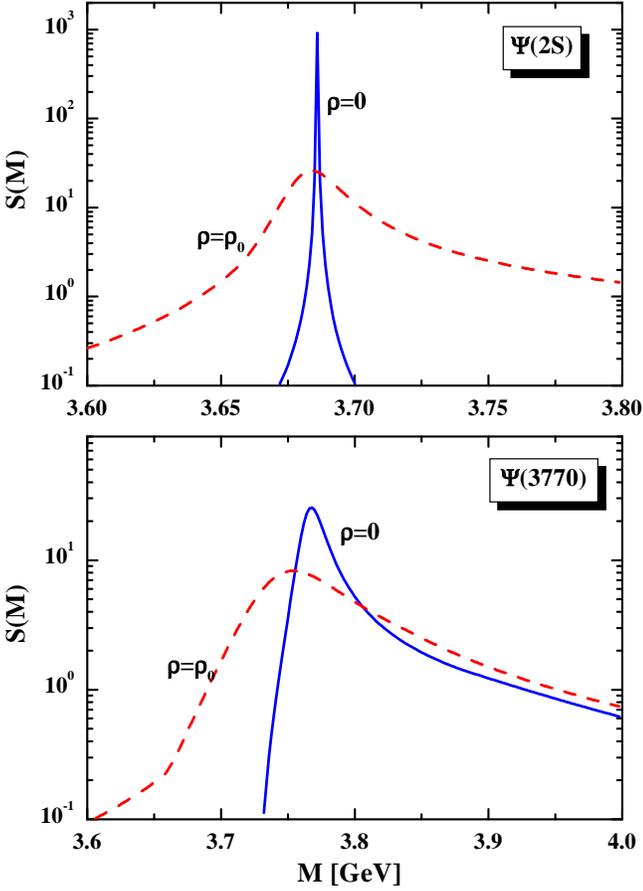,width=8.5cm}}
 \caption{The spectral functions $S(M)$ for the $\Psi(2S)$ (upper part)
 and $\Psi(3770)$ (lower part)
in 'free' space (solid lines) and at density $\rho_0$ (dashed
lines) according to the model described in the text.}
 \label{bild1}
\end{figure}

In the nuclear medium the $D,\bar{D}$ mesons are expected to show
attractive mass shifts approximately linear in the density
$\rho_A$ according to the QCD sum rule studies in Ref.
\cite{Haya}. We thus adopt
\begin{equation}
\label{drop} \Delta M_{D/\bar{D}} \approx - \alpha_{D/\bar{D}}(p)
\frac{\rho_A}{\rho_0}
\end{equation}
where $p$ denotes the meson momentum relative to the nuclear
target at rest and $\rho_0 \approx$ 0.168  fm$^{-3}$ the nuclear
matter density. The coefficients $\alpha_{D/\bar{D}}(p)$ should be
determined by a microscopic theory, however, we will first
approximate them by the constants $\alpha_{D} = 0.06$ MeV and
$\alpha_{\bar{D}} = 0.01$ MeV, respectively, following the
suggestions in Ref. \cite{Haya}. Since an explicit momentum
dependence of these coefficients is expected to reduce the
self-energy effects for high momenta  this particular
model should lead to upper estimates for the medium effects.

With the in-medium masses for $D,\bar{D}$ fixed by (\ref{drop}) we
can calculate the in-medium decay widths of the charm vector
mesons (\ref{channel}) and via (\ref{coll}) the total decay width
(\ref{gamma}) assuming a total $VN$ cross section of 6 mb, which
is extracted from the $J/\Psi$ absorption data of the NA50
Collaboration \cite{NA50}. Note that photoproduction of $J/\Psi$'s
on nucleons give a total $J/\Psi-N$ cross section of 3-4 mb
\cite{Huefner}, however, the $\Psi(2S)$ and $\Psi(3770)$ cross
sections with nucleons should be larger by about a factor of 2 due
to their larger radial size. The resulting spectral functions
$S(M)=2 M A(M)$ -- including again the electromagnetic and light
hadron  decay widths -- are shown in Fig. 1 for the $\Psi(2S)$ (upper part) and $
\Psi(3770)$ (lower part)  for nuclear matter
density $\rho_0$  by the dashed lines. We observe substantially
broadened spectral functions for the $\Psi(2S)$ as well as
$\Psi(3770)$ due to collisional broadening ($\approx$ 15 MeV) and
a larger phase space for $D+\bar{D}$ decay. Note, that the
threshold for $D+\bar{D}$ decay according to (\ref{drop}) at
density $\rho_0$ moves down to 3.668 GeV -- which is 18 MeV below
the $\Psi(2S)$ pole mass of 3.686 GeV -- such that a larger
fraction of the $\Psi(2S)$ spectral distribution can decay to
$D+\bar{D}$ in the medium. The life time of this resonance
decreases accordingly from 656 fm/c in free space to $\sim 9$ fm/c
at density $\rho_0$ such that some fraction of the $\Psi(2S)$ is
expected to decay inside heavy nuclei such as $^{208}Pb$.

\subsection{Production cross sections in $\bar{p} p$ and $\bar{p} A$ reactions}
 The mass differential production of the vector mesons in $\bar{p} p$ reactions
can be described by Breit-Wigner resonance formation on the basis
of the spectral function $A(M^2)$ (\ref{BW}) - employing the
proper kinematics - provided that the branching of $\bar{p}p
\rightarrow V$ is known,
\begin{equation}
 \sigma_V(M) = \frac{3}{4} \  Br_{p\bar{p}
\rightarrow V}\  4 M A_V(M) \ \Gamma^{tot}_V(M) \
\frac{\pi^2}{k^2}, \label{cross}
\end{equation}
where the factor 3/4 stems from the ratio of spin factors and $k$
denotes the momentum of the $p$ (or $\bar{p}$) in the cms.
Following our previous works \cite{Cass00,Golub02} we adopt
$Br(p\bar{p} \rightarrow \Psi(3770)) \approx 2 \times 10^{-4}$
which is very similar to the branching of the $\Psi(2S)$ state
\cite{PDG}. The resulting cross sections for the mesons $V$ in
$\bar{p} p$ reactions are displayed in Fig. 2 (upper part) as a
function of the antiproton kinetic energy $T_{\bar{p}} (=T)$ in the
laboratory reflecting the resonance structure (\ref{BW}) and the
kinematics from (\ref{cross}) in free space. Note, that due to the
small width of the $\Psi(2S)$ ($\Gamma_{tot} \approx$ 0.3 MeV) the
kinematics have to match the resonance conditions very well.

\begin{figure}[t]
\centerline{\psfig{figure=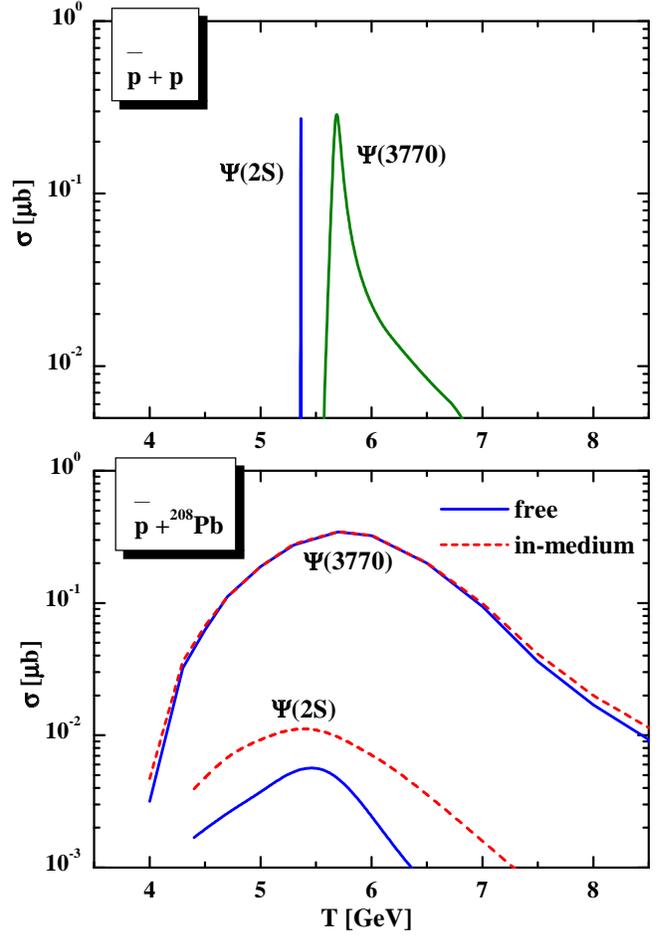,width=8.5cm}}
 \caption{The calculated cross sections for $\Psi(2S)$ and
$\Psi(3770)$ for $\bar{p}p$ (upper part) and $\bar{p}+^{208}Pb$
reactions (lower part) as a function of the antiproton kinetic
energy $T$ in the laboratory employing 'free' (solid lines) and
'in-medium' spectral functions (dashed lines) as described in the
text. Note, that no rescattering effects are included in these
primary formation cross sections.}
 \label{bild2}
\end{figure}

In order to simulate events for the reaction $\bar{p}A \rightarrow
V $ we use the Multiple Scattering Monte Carlo (MSMC) approach. An
earlier version of this approach -- denoted as Intra-Nuclear
Cascade (INC) model -- has been applied to the analysis of $\eta$
and $\omega$ production in $\bar p A$ and $p A$ interactions in
Refs.~\cite{Golub92,Golub93,Golub96,Golub97}. The production of
the hidden charm vector resonances on nuclei then can be evaluated
within the MSMC by propagating the antiproton in the nucleus and
calculating the elementary $\bar{p} p \rightarrow V$ production
cross section  at the annihilation point, where the local momentum
distribution is taken into account in the local Thomas-Fermi
approximation.

The numerical results for the formation cross sections of the
resonances $\Psi(2S)$ and $\Psi(3770)$ on $^{208}Pb$ are shown in
the lower part of Fig. 2 as a function of $T_{\bar{p}}$ -- using
the 'free' spectral functions (solid lines) -- indicating that the
resonance formation is further smeared out by Fermi motion and the
maximum in the cross section becomes much reduced in case of the
$\Psi(2S)$ inspite of the larger $\bar{p}$ annihilation cross
section on $^{208}Pb$. As discussed in more detail in Ref.
\cite{Golub02} the averaging over the Fermi motion in the target
leads to kinematical conditions that do not match the resonance
properties in case of a small total width and thus lead to a
suppression of resonance formation. For the $\Psi(3770)$ and its
width of $\sim$ 25 MeV the Fermi averaging is less crucial and
compensated at $T_{\bar{p}}$ = 5.7 GeV by the larger geometrical
size of the $Pb$ target.

The formation cross section modifies only slightly for the
$\Psi(3770)$ (dashed line) when employing the density dependent
in-medium spectral function as described above. Thus the
excitation function in the antiproton (lab.) kinetic energy
$T_{\bar{p}}$ does not provide relevant
information on the medium modifications of the $\Psi(3770)$ due to
the dominance of Fermi motion in the nuclear target. The situation
is different for the $\Psi(2S)$ since here a large increase in the
total width is expected in the medium (cf. Fig. 1). As a
consequence the relative suppression by Fermi smearing becomes
less pronounced and we end up with a sizeably enhanced cross
section on nuclei for the in-medium case. Note, however, that
$\Psi(2S)-N$ dissociation reactions will reduce this enhancement
again as well as in-medium $D+\bar{D}$ decays such that the net
effect to be seen in the $e^+e^-$ channel is even a $\sim$ 40\%
reduction of the formation cross section relative to a calculation
with a 'free' spectral function (see below).

\subsection{Vector meson propagation and decay}
Necessary parameters for a Monte Carlo (MC) simulation of
rescattering are the elastic and inelastic $V N$ scattering cross
sections and slope parameters $b$ for the differential elastic
cross sections $d\sigma_{el}/dt$, which are approximated by
\begin{equation} \frac{d \sigma_{el}}{dt} \sim \exp(bt), \end{equation}
where $t$ is the momentum transfer squared. These parameters as
well as the masses of the rescattered particles determine the
momentum and angular distributions of the particles in the final
state. As in our previous studies \cite{Cass00,Golub02} we use
$b$= 1 GeV$^{-2}$ for $D,\bar{D}+N$ and $VN$ scattering as an
educated guess.

Furthermore, the inelastic cross sections of the vector mesons
with nucleons have to be specified. Since the relative momenta in
the $V-N$ system for rescattering are in the order of a few GeV
one might use high energy geometric cross sections for the
dissociation to open charm mesons. For  the two resonances studied
here we adopt an inelastic cross section of 6 mb as well as
$\sigma_{el}$ = 1 mb. As mentioned above the inelastic cross
section of 6 $mb$ is larger by about a factor of $\sim 2$ in
comparison to the $J/\Psi - N$ cross section  due to the larger
size of the radially excited states.

The resonances are propagated with their actual mass $M$ -
selected by MC according to the in-medium spectral function
(\ref{BW}) - with velocity ${p}_V/E_V$ and decay in time according
to the differential equation
\begin{equation} \frac{d P_V(M)}{dt} = - \frac{1}{\gamma} \  \Gamma^V_{tot}(M,\rho_A)
\ P_V(t)
\end{equation}
with the total width (\ref{gamma}), while $\gamma$ denotes the
Lorentz $\gamma$-factor. Their decay to dileptons or
 $D\bar{D}$ is, furthermore, determined
by Monte-Carlo according to the mass differential branching ratios
at the local density $\rho_A$. The off-shell propagation of the
charmonia and evolution of their spectral functions is performed
in line with the off-shell transport equations developed in Refs.
\cite{Caju}.

The open charm mesons from the decaying vector mesons in the
present case have a light quark $q=(u,d)$ content apart from the
$\bar{c}$ or $c$ quark. In the constituent quark model we get
${D}^+ = (c\bar{d})$, ${D}^- = (\bar{c} {d})$, $\bar{D}^0 =
(\bar{c}{u})$, ${D}^0 = (c\bar{u})$ and the same composition for
the related vector states. In order to estimate their elastic and
inelastic cross section with nucleons we use the analogy to $KN$
and $\bar{K}N$ interactions. It is well established experimentally
from $K^+ p$ and $K^-p$ reactions that light quark exchanges with
nucleons ($(uud)$ or $(udd)$) have different strength. Whereas the
$K^+ (u \bar{s})$ scatters only elastically with nucleons at low
momentum - since the $\bar{s}$ cannot be exchanged with a light
quark of a nucleon - the $K^- (s\bar{u})$ cross section is
dominated by resonant $s$-quark transfer leading to strange
baryons such as $\Lambda$ or $\Sigma$ hyperons. Similar relations
are expected to hold for the $D, \bar{D}$ mesons \cite{Alex99},
where especially the light quark contribution should give much
larger cross section on nucleons than the $c\bar{c}$ vector
resonances. This analogy is based on $SU(4)_{flavor}$ symmetry
\cite{Lin,Ko00,Zhang2} which, however, might be broken
substantially in view of the different geome\-trical sizes of $K$-
and $D$-mesons. At present this is an open question which has to
be settled by experiment. Corresponding experiments have been
proposed in Ref. \cite{Golub02}.

For our calculations  we adopt the following cross sections -
taken as constants in the momentum regime of interest - $$
\sigma^{el}_{{D} N} = \sigma^{el}_{\bar{D} N} = 10 \
 mb; \ $$ \begin{equation} \sigma^{inel}_{\bar{D} N} \approx 0; \
\sigma^{inel}_{{D} N} = 10 \ mb .  \label{parameters}
\end{equation}
Here the inelastic cross sections of $D$-mesons refer to $c$-quark
exchange reactions with nucleons. Furthermore, charge ex\-change
reactions like ${D}^+ n \leftrightarrow {D}^0 p$ or ${D}^- p
\leftrightarrow \bar{D}^0 n$ are incorporated with a constant
cross section $\sigma_{\rm{q exc.}}$ = 12 $mb$ (cf. Ref.
\cite{Golub02}).

\section{Resonance production in $\bar{p} A$ reactions}
Within the model described in Section 2 we  now can calculate the
cross sections for the $c \bar{c}$ resonances in $\bar{p} A$
reactions for all targets of interest and obtain the information
about in-medium resonance decays (for densities $\rho_A \geq 0.03
fm^{-3}$) or vacuum decays, respectively. We note that apart from
the resonant production of open charm mesons they may also be
produced directly as $(c\bar{q})$ and ($\bar{c}q$) pairs in
$\bar{p}N$ annihilation. This channel strongly dominates in
annihilation reactions on neutrons in the nucleus since the
charmonium production is forbidden by charge conservation at low
energy above threshold. At higher invariant energies a pion might
balance the charge in the $\Psi + \pi$ final channel. For our
estimates we have employed the Regge-model analysis of Ref.
\cite{Kaidalov} for the elementary $\bar{p} N \rightarrow D,
\bar{D}$ cross section as also used by Sibirtsev et al. in Refs.
\cite{Alex99,Sibirtsev}. As already shown in Ref. \cite{Golub02}
the nonresonant production of $D,\bar{D}$ pairs at $T_{lab}$ = 5.7
GeV is expected to be about the same as the resonant production
via $\Psi(3770)$ excitation and decay in case of heavy nuclei.

\subsection{Open charm propagation and rescattering}
The open charm mesons produced by resonance decay or nonresonant
channels  rescatter in the nuclear medium either elastically or
inelastically. Furthermore, they change their quasi-particle
properties (or mass) dynamically according to (6) while propagating from the
nuclear medium to the vacuum. This dynamical problem can be well
addressed by the (MSMC) approach as well for all bombarding
energies and targets.

\begin{figure}[t]
\centerline{\psfig{figure=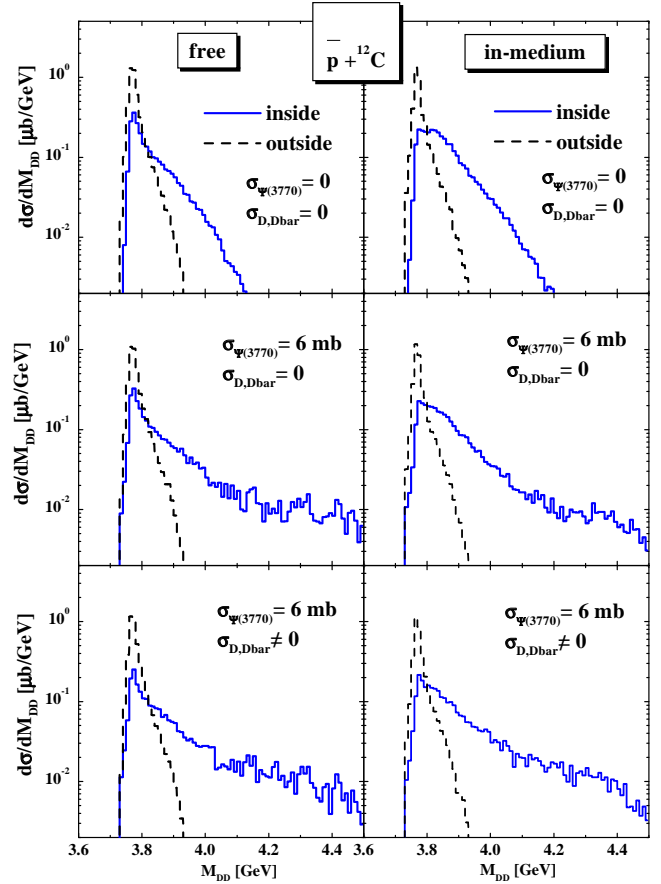,width=8.5cm}}
 \caption{The calculated $D+\bar{D}$ invariant mass
distributions for $\bar{p}+^{12}C$ reactions at $T_{\bar{p}}$ =
5.7 GeV within various limits for the 'in-medium' decay (solid
histograms) and 'vacuum' decays (dashed histograms), respectively.
The diagrams on the l.h.s. correspond to calculations with the
'free' spectral function for the $\Psi(3770)$ whereas the diagrams
on the r.h.s. stem from calculations with the 'in-medium' spectral
functions. In the upper plots the rescatterings of the
$\Psi(3770)$ and $D,\bar{D}$-mesons have been neglected whereas in
the middle plots $\Psi(3770) + N$ dissociation has been included.
The lower plots additionally include the $D, \bar{D}$ elastic and
inelastic rescattering processes as specified in Section 2.}
 \label{bild3}
\end{figure}

In order to properly interpret the final results it is instructive
to show the $D+\bar{D}$ invariant mass spectra within different
dynamical limits.  To this aim we display in Figs. 3 and 4 (upper
left parts) the invariant mass spectrum of $D+\bar{D}$ mesons from
$\Psi(3770)$ decays in $\bar{p} + ^{12}C$ and $\bar{p} + ^{208}Pb$
reactions at $T_{\bar{p}}$ = 5.7 GeV using the 'free' charmonium
spectral function and branching ratios while discarding open charm
and $\Psi(3770)$ rescattering. As seen from Fig. 3 and Fig. 4
(upper left parts) the in-medium decays to $D+\bar{D}$ (solid
histograms) make up the major part of the invariant mass
distribution at masses above 3.85 GeV since the life time of high
mass $\Psi(3770)$ becomes very short in nuclei.  According to the
same argument the life time of $\Psi(3770)$'s at low invariant
masses is rather high due to the very limited phase space for
$D+\bar{D}$ decay, such that we obtain a preferential decay in the
vacuum ($\rho_A \leq 0.03  \ fm^{-3}$) for invariant masses below
3.8 GeV especially in case of the light $^{12}C$-target.

For the next calculations we switch on the $\Psi(3770)-N$
rescattering, however, discard  $D,\bar{D}$ mass shifts and
rescattering (middle left parts in Figs. 3 and 4). The effects of
$\Psi(3770)$ rescatterings are esssentially a broadening of the
$D+\bar{D}$ invariant mass distribution to high invariant masses
and a reduction of the vacuum contribution (dashed histograms) due
to dissociation reactions.  This rather obvious medium
modification is enhanced when including $D,\bar{D}$ rescattering
(lower left parts in Figs. 3 and 4) which additionally leads to a
depletion of the invariant mass distribution in the resonance peak
since rescattering changes the invariant mass of the $D \bar{D}$
pair observed asymptotically. These quite drastic medium
modifications compete with the spectral changes of the
$\Psi(3770)$ in the medium due to the $D,\bar{D}$ potentials.

\begin{figure}[t]
\centerline{\psfig{figure=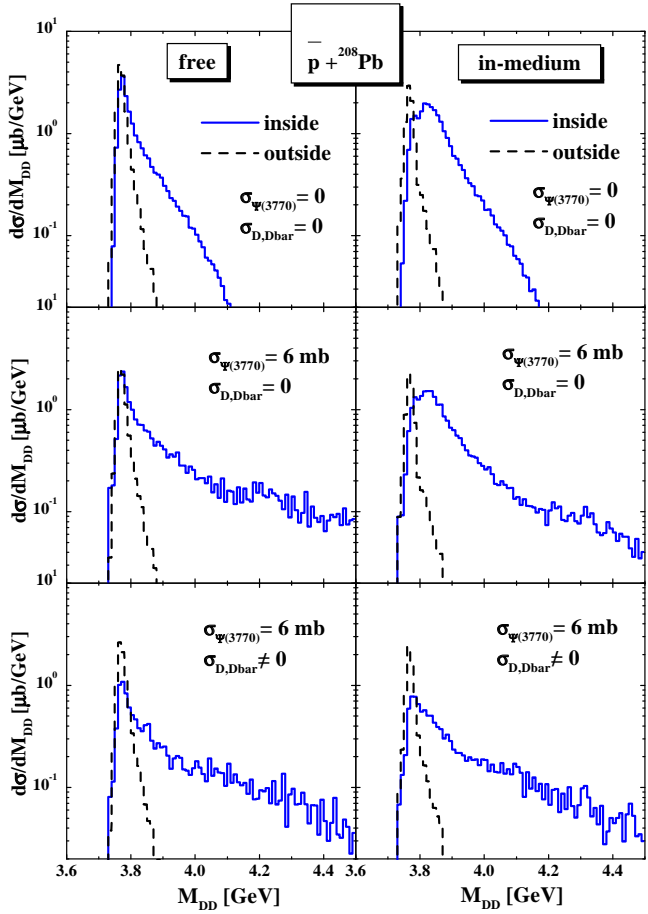,width=8.5cm}}
 \caption{The calculated $D+\bar{D}$ invariant mass
distributions for $\bar{p}+^{208}Pb$ reactions at $T_{\bar{p}}$ =
5.7 GeV within various limits for the 'in-medium' decay (solid
histograms) and 'vacuum' decays (dashed histograms), respectively.
The various calculations correspond to the same limits as in Fig.
3.}
 \label{bild4}
\end{figure}

In the r.h.s. of Figs. 3 and 4 we show the corresponding
$D+\bar{D}$ invariant mass distributions for the dynamical
$\Psi(3770)$ spectral function without rescatterings (upper right
parts), with $\Psi(3770)$ rescattering (middle right parts) and
also including $D, \bar{D}$ interactions with nucleons (lower
right parts). Whereas the high mass spectra differ substantially
compared to the 'free' $\Psi(3770)$ spectral function (l.h.s.)
when discarding final state interactions, the final mass spectra
-- including $\Psi(3770)-N$ and $D, \bar{D}$ interactions -- look
very similar. One might have expected an enhancement in the low
mass region below 3.77 GeV, however, the in-medium decays to
$D+\bar{D}$ at lower invariant masses finally show up in the
vacuum at higher invariant masses (above the $D+\bar{D}$ vacuum
threshold) since the open charm mesons regain their 'free' masses
when moving out of the medium.

A direct comparison of the $D+\bar{D}$ invariant mass
distributions -- summing up the 'inside' and 'outside' components
-- is displayed in Fig. 5 for $^{12}C$ and $^{208}Pb$ for the case
of the 'free' spectral function (solid histograms) and the
in-medium spectral functions (dashed histograms). Here all final
state interactions have been included in the calculations. Within
statistics there is practically no difference between the spectra
for the $Pb$ target such that $D+\bar{D}$ invariant mass
spectroscopy does not qualify much to probe the in-medium
$\Psi(3770)$ spectral function. This becomes even more apparent
when comparing additionally to the background from nonresonant
$D\bar{D}$ production (dotted lines 'BG') which -- integrated over
invariant mass -- is expected to be of comparable magnitude in
cross section at $T_{\bar{p}}$ = 5.7 GeV \cite{Golub02}. Thus,
apart from the vicinity of the resonance pole, the nonresonant
background will dominate the high mass region of the invariant
mass spectra. We mention that in order to calculate the invariant
mass distribution from the nonresonant background we have assumed
2-body $S$-wave kinematics for the process $\bar{p} p \rightarrow
D\bar{D}$.

\begin{figure}[t]
\centerline{\psfig{figure=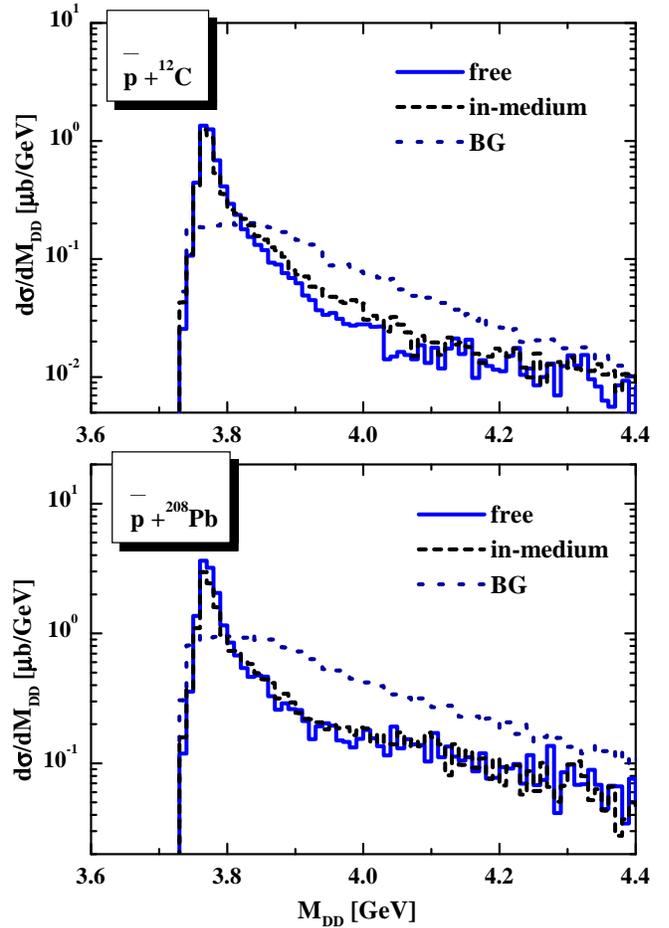,width=8.5cm}}
 \caption{The calculated $D+\bar{D}$ invariant mass
distributions for $\bar{p}+^{12}C$ and $\bar{p}+^{208}Pb$
reactions at $T_{\bar{p}}$ = 5.7 GeV as obtained from summing up
the 'inside' and 'outside' decay contributions from Figs. 3 and 4
where all final state interactions have been included. The solid
histograms correspond to the case of the 'free' $\Psi(3770)$
spectral function while the results for the 'in-medium' spectral
functions are shown in terms of the dashed histograms. The dotted
lines (BG) show additionally the expected background from
nonresonant $D, \bar{D}$ production as calculated by 2-body final
state kinematics.}
 \label{bild5}
\end{figure}

Without explicit representation we mention that the perspectives
to study the in-medium properties of the $\Psi(2S)$ by $D+\bar{D}$
invariant mass spectroscopy  at $T_{\bar{p}}$ = 5.4 GeV are rather
discouraging, since the contribution from the $\Psi(3770)$ decays
by far exceeds the signal from the $\Psi(2S)$ decays (cf. Fig.
\ref{bild2}) and the in-medium $D, \bar{D}$ move on-shell again,
such that only a tiny contribution above threshold survives.
According to the authors point of view such tiny effects will be
extremely hard to identify experimentally.

However, it remains to be seen if the expected in-medium
modifications of the charmonia might be detected via the dilepton
decay mode since the $e^+e^-$ (or $\mu ^+ \mu^-$) pairs do not
suffer from strong final state interactions and also have no
threshold in the vicinity of the charmonium pole masses.

\subsection{Resonant dilepton decays in $\bar{p} A$ reactions}
We continue with the $e^+ e^-$ decay modes of the charmonia that
have a decay width of $\sim$ 2.2 keV for the $\Psi(2S)$ and $\sim$
0.26 keV for the $\Psi(3770)$, respectively \cite{PDG}. Inspite of
these low decay widths the dilepton decay mode might provide
essential information about the in-medium properties of the
charmonia.

In our actual calculations we assume the dileptonic decay width to
be constant as a function of invariant mass $M_{e^+e^-}$ since in
case of the charmonia the total width is very small compared to
the pole masses. We start with $\Psi(3770)$ resonant production in
$\bar{p}+ ^{12}C$ and $\bar{p}+ ^{208}Pb$ reactions at
$T_{\bar{p}}$ = 5.7 GeV and consider the same limits as for the
$D+\bar{D}$ decay mode. The numerical invariant mass spectra for
these reactions are shown in Figs. 6 and 7, respectively.

\begin{figure}[t]
\centerline{\psfig{figure=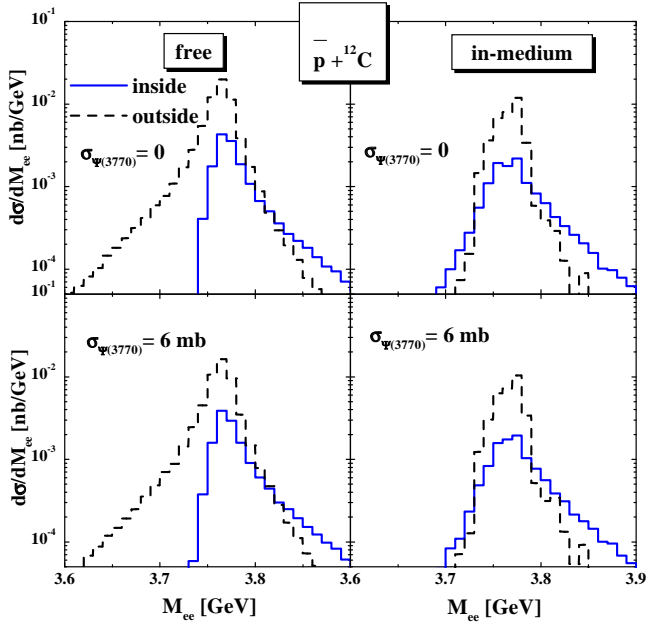,width=8.5cm}}
 \caption{The calculated $e^+ e^-$ invariant mass
distributions for $\bar{p}+^{12}C$ reactions at $T_{\bar{p}}$ =
5.7 GeV within various limits for the 'in-medium' decays (solid
histograms) and 'vacuum' decays (dashed histograms), respectively.
The diagrams on the l.h.s. correspond to calculations with the
'free' spectral function for the $\Psi(3770)$ whereas the diagrams
on the r.h.s. stem from calculations with the 'in-medium' spectral
functions. In the upper plots the rescatterings of the
$\Psi(3770)$ mesons have been neglected whereas in the lower plots
$\Psi(3770) + N$ dissociation has been included.}
 \label{bild6}
\end{figure}

As seen from Figs. 6 and 7  the in-medium dilepton decays (40\%
for $Pb$) dominate for high invariant mass pairs whereas the low
mass components -- below the $D+\bar{D}$ threshold -- only stem
from vacuum decays (60\% for $Pb$) in case of the 'free'
$\Psi(3770)$ spectral functions (l.h.s.). This phenomenon is
easily attributed to the long life time of the $\Psi(3770)$ for
low invariant mass such that it only decays in vacuum and
'radiates' dileptons for a long time. When including
$\Psi(3770)-N$ dissociation reactions (lower parts of Figs. 6 and
7) the shape and relative contribution from 'inside' and 'vacuum'
decays does not change very much, only the magnitude of the mass
differential cross section is reduced for $Pb$  by a factor of
$\sim$0.63. For the 'in-medium' $\Psi(3770)$ spectral function the
individual results look different (r.h.s.), since now the
resonance can decay to $D \bar{D}$ in the medium at lower
invariant masses, too, and the life time for low mass $\Psi(3770)$
decreases substantially. Accordingly, there is also a large
'inside' fraction of dilepton decays for lower masses which now no
longer can decay in the vacuum. The result is a narrowing of the
$\Psi(3770)$ spectral function in the 'outside' contribution from
the lower mass side. The relative contributions from 'inside' and
'outside' decays do not change substantially when including again
$\Psi(3770)+N$ dissociation reactions (lower parts of Figs. 6 and
7). Only the magnitude of the differential cross section becomes
suppressed as in case of the 'free' spectral function (l.h.s.).

\begin{figure}[t]
\centerline{\psfig{figure=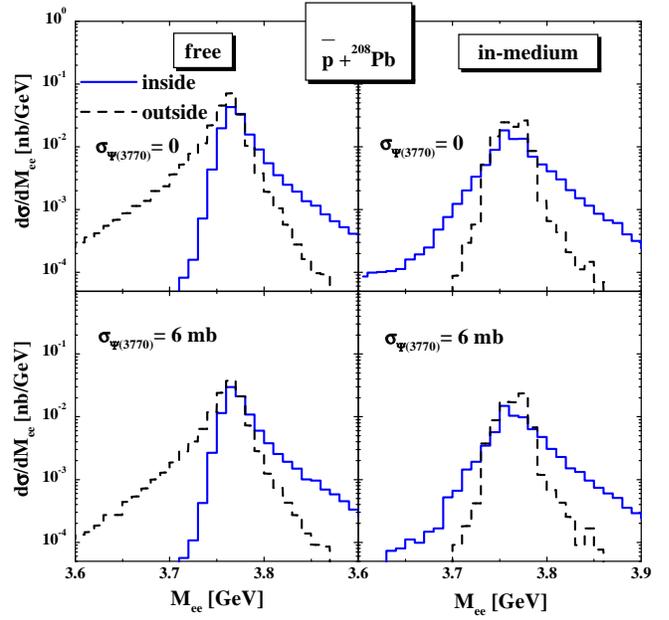,width=8.5cm}}
 \caption{The calculated $e^+ e^-$ invariant mass
distributions for $\bar{p}+^{208}Pb$ reactions at $T_{\bar{p}}$ =
5.7 GeV within various limits for the 'in-medium' decay (solid
histograms) and 'vacuum' decays (dashed histograms), respectively.
The calculations correspond to the same limits as in Fig. 6
for the $^{12}C$ target.}
 \label{bild7}
\end{figure}

The net spectra (possibly) to be  seen experimentally are
displayed in Fig. 8 for both targets in case of the 'free' (solid
histograms) and 'in-medium' $\Psi(3770)$ spectral functions
(dashed histograms) where the 'inside' and 'outside' components
have been added up. Apart from the fact, that the differential
cross sections are below 0.02 and 0.1 nb/GeV, respectively, the
me\-dium modifications only show up in a suppression  of the low
mass distributions since the high mass tails are practically the
same. Thus the net effect is not an 'enhancement' of lower mass
dileptons as in case of the $\rho$ and $\omega$ mesons
\cite{Cass99,Rapp}, but a relative 'suppression'. If this
suppression could be seen in experiment it should provide valuable
information on the $D, \bar{D}$ properties in the nuclear medium.

\begin{figure}[t]
\centerline{\psfig{figure=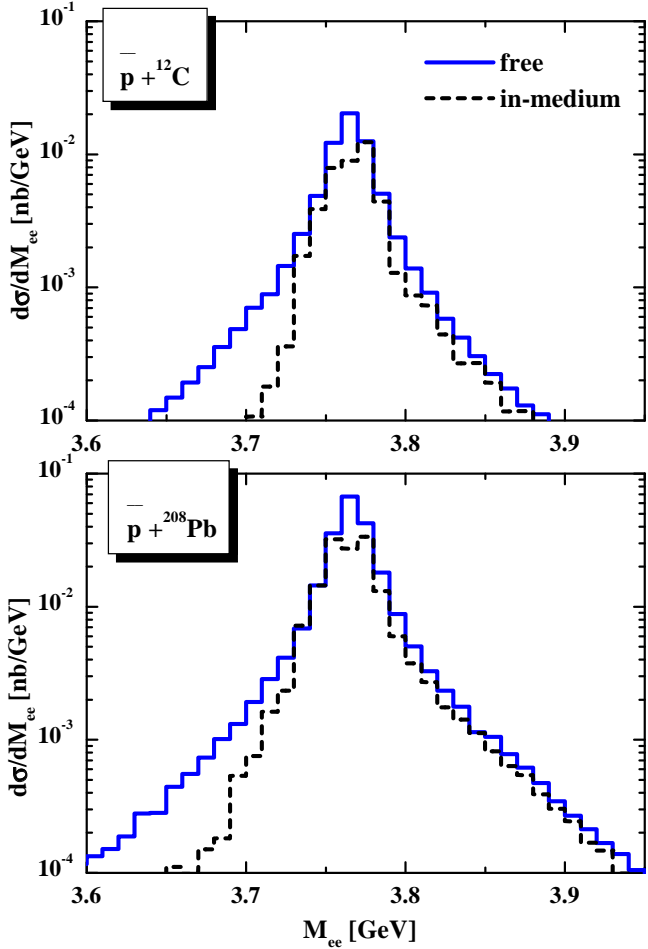,width=8.5cm}}
 \caption{The calculated $e^+ e^-$ invariant mass
distributions for $\bar{p}+^{12}C$ and $\bar{p}+^{208}Pb$
reactions at $T_{\bar{p}}$ = 5.7 GeV as obtained from summing up
the 'inside' and 'outside' decay contributions from Figs. 6 and 7
where all final state interactions have been included. The solid
histograms correspond to the case of the 'free' $\Psi(3770)$
spectral function while the results for the in-medium spectral
functions are shown in terms of the dashed histograms. }
 \label{bild8}
\end{figure}

The question remains, to which extent a substantial broadening of
the $\Psi(2S)$ in the medium will leave its traces in the dilepton
invariant mass spectra. Our actual calculations for the system
$\bar{p}+^{12}C$ at $T_{\bar{p}}$ = 5.5 GeV show no effects within
statistics since the average density in $^{12}C$ is substantially
below nuclear matter density and the life time of the $\Psi(2S)$
is large (in fm/c) compared to the size of this nucleus (in fm).
For $\bar{p} + ^{208}Pb$ at $T_{\bar{p}}$ = 5.5 GeV there is a
small fraction of $\Psi(2S)$ (about 2\%) that decay to $e^+e^-$
inside the nucleus (dashed histogram in Fig. 9) and show a
spectral distribution that is compatible with a Breit-Wigner shape
of 6 MeV width (cf. thin solid line in Fig. 9), which is larger by
a factor of 20 than the 'free' width of 0.3 MeV. However, when summing
up the 'inside' and 'outside' dilepton decays again only a small
broadening in the tails of the invariant mass spectra survives
(solid histogram in Fig. 9). We note, that this effect might be
hidden in the background from Drell-Yan dileptons and hard to be
seen experimentally.

\begin{figure}[t]
\centerline{\psfig{figure=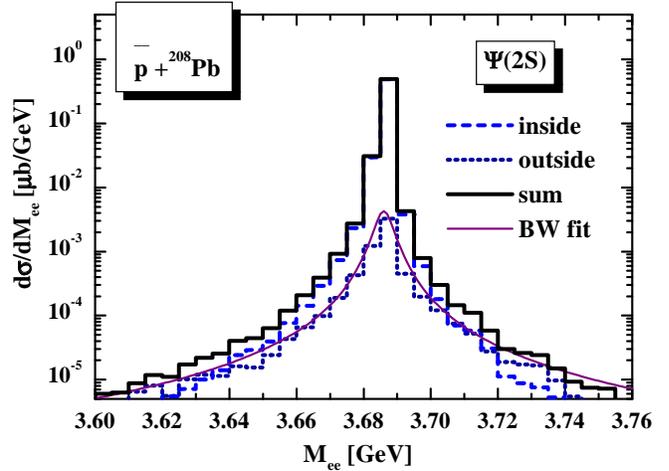,width=8.5cm}}
 \caption{The calculated $e^+ e^-$ invariant mass
distribution for  $\bar{p}+$ $^{208}Pb$ reactions at $T_{\bar{p}}$ =
5.5 GeV from the 'inside' (dashed histogram)  and 'outside' decay
contribution (dotted histogram) employing the in-medium $\Psi(2S)$
spectral function described in the text. The 'inside' component
can be fitted by a Breit Wigner distribution with a width of 6 MeV
(thin solid line). The total spectrum is given by the solid
histogram.}
 \label{bild9}
\end{figure}

\section{Summary} In this study we have explored the perspectives of
measuring the dynamics of hidden charm vector mesons and open
charm mesons in antiproton induced reactions on nuclei by
$D+\bar{D}$ and $e^+e^-$ invariant mass spectroscopy. Such
experimental studies will be possible at the high-energy
antiproton storage ring (HESR) proposed for the future GSI
facility \cite{GSIfut}. The Multiple Scattering Monte Carlo (MSMC)
calculations have been based on the resonance production concept
with resonance properties from the PDG \cite{PDG} in vacuum,
however,  modified in the nuclear medium according to phase space
(cf. Section 2) involving dropping masses for the $D$ and
$\bar{D}$ mesons (\ref{drop}). The actual values have been taken
from the QCD sum rule studies of Ref. \cite{Haya}.

We find that for a heavy nucleus like $^{208}Pb$ the fraction of
'inside' decays to dileptons or $D+\bar{D}$ reaches $\sim$ 40\%
for the $\Psi(3770)$, which at first sight looks promising.
However, the in-medium effects in the $D+\bar{D}$ invariant mass
spectra are hard to see experimentally since for invariant masses
from the $D+\bar{D}$ threshold to the $\Psi(3770)$ pole mass the
final spectrum is only slightly modified while at high invariant
masses it is covered by the nonresonant background (Fig. 5). The
$e^+e^-$ invariant mass spectra from the in-medium $\Psi(3770)$
show a decrease at small masses (below the $D+\bar{D}$ threshold)
since these charmonia preferentially now decay via the $D+\bar{D}$
channel in the nucleus. Thus the net effect is not an
'enhancement' of lower mass dileptons as in case of the $\rho$ and
$\omega$ mesons \cite{Cass99,Rapp}, but a relative 'suppression'.
If this suppression could be seen in experiment it might provide
valuable information on the $D, \bar{D}$ properties in the nuclear
medium.

Furthermore, we find that for the in-medium effects incorporated
the $D,\bar{D}$ decays of the $\Psi(2S)$ in the nucleus -- that
become possible due to a dropping of the $D, \bar{D}$ masses --
are unlikely to be seen due to the much larger yield from the
$\Psi(3770)$ at the same bombarding energy (Fig. 2). On the other
hand, a small net broadening of the $\Psi(2S)$ resonance tails
survives in the dilepton channel according to our calculations
(Fig. 9) which, however, will be hard to separate from the
Drell-Yan background.

We close in noting that the in-medium effects explored in this
work are based on QCD sum rule studies for hidden charm and open
charm mesons at vanishing relative momentum and low nuclear
densities \cite{Klingl,Haya}. In the $\bar{p} + A$ reactions
investigated  the mesons  actually have momenta of 2-4 GeV/c
\cite{Golub02} with respect to the target at rest, such that the
in-medium effects explored more likely correspond to an educated
guess. If the $D, \bar{D}$ potentials (\ref{drop}) decrease
substantially in magnitude with momentum then all effects pointed
out here will decrease accordingly and practically no medium
modifications should be seen experimentally. On the other hand,
the open charm potentials (\ref{drop}) might be even larger in
nature and thus the effects become enhanced. Nevertheless, an
answer here has to come from related experimental studies e.g. at
the HESR \cite{GSIfut} which appear sufficiently promising.

\section*{Acknowledgements}
This work was supported by DFG under grants No. 436 RUS 113/600,
CA 124/4 and RFFI.

\end{document}